\documentclass[%
reprint,
 amsmath,amssymb,
 aps,
]{revtex4-2}

\usepackage{graphicx}
\usepackage{dcolumn}
\usepackage{bm}
\RequirePackage[colorlinks,citecolor=blue,urlcolor=blue,linkcolor=blue]{hyperref}
\usepackage[caption=false]{subfig}
\usepackage{verbatim}
\usepackage{physics}
\usepackage{float}
\usepackage{graphicx}
\usepackage{dcolumn}
\usepackage{bm}
\usepackage{booktabs}
\usepackage[table,xcdraw]{xcolor}
\usepackage{multirow}
\usepackage{array}
\usepackage[titletoc]{appendix}
\allowdisplaybreaks

\newcolumntype{P}[1]{>{\centering\arraybackslash}p{#1}}

\begin{document}

\preprint{APS/123-QED}

\title{Properties of hot finite nuclei and associated correlations with infinite nuclear matter}

\author{Vishal Parmar}
\author{Manoj K Sharma}%
 \affiliation{School of Physics and Materials Science, Thapar
    Institute of Engineering and Technology, Patiala-147004, India.}

    \email{physics.vishal01@gmail.com}
 \author{S K Patra}
\affiliation{%
Institute of Physics, Bhubaneswar 751005, India \\Homi Bhaba National Institute, Training School Complex, Anushakti Nagar, Mumbai 400 085, India}%

\date{\today}

\begin{abstract}

This work aim to study the various thermal characteristics of nuclei in view of the saturation and critical behavior of infinite nuclear matter. The free energy of a nucleus is parametrized using the density and temperature-dependent liquid-drop model and interaction among nucleons is worked out within the effective relativistic mean-field theory (E-RMF).  The effective mass (m,$^*$) and critical temperature of infinite symmetric nuclear matter ($T_c$) of a given E-RMF parameter force play a seminal role in the estimation of thermal properties.   A larger (m$^*$) and  $T_c$ of the E-RMF set estimate larger excitation energy, level density, and limiting temperature $(T_l)$ for a given nucleus.  The limiting temperature of a nucleus also depends on the behavior of the nuclear gas surrounding the nucleus, making the equation of state (EoS) at subsaturation densities an important input.  A stiff  EoS in the subsaturation region estimates a higher pressure of the nuclear gas making it less stable.  Since the $T_c$ plays an important part in these calculations, we perform a Pearson correlation statistical study of fifteen E-RMF parameter sets, satisfying the relevant constraint on EoS. Effective mass seems to govern the thermal characteristics of infinite as well as finite nuclear matter in the framework of E-RMF theory.

\end{abstract}

\maketitle


\section{\label{introduction} Introduction}  

One of the  astonishing universality in the laws of nature is the resemblance between the nuclear and the molecular force. Molecular force is of van der Waals type and nuclear force behaves similarly, albeit on the different energy scale.  Therefore one may arrive at the notion that nuclear matter should undergo a liquid-gas phase transition (LGPT) like a classical liquid drop. This phenomenon of LGPT in both infinite nuclear matter and finite nuclei is an important feature of heavy-ion-induced reactions (HIR) \cite{PhysRevLett.72.3321, tlimexpdata, WU1997385}. In these reactions, the participating hot nuclei undergo multi-fragmentation after the initial dynamic stage of compression upon reaching sub-saturation density ($\approx 0.2\rho_0$) \cite{KARNAUKHOV200691}. In this sub-saturation density region, the properties of nuclei are modified \cite{PhysRevLett.94.162701, PhysRevC.95.061601, BORDERIE201982} which are very essential for the understanding of thermodynamics of hot nuclei, and the medium in which they are created. The knowledge of the nuclear matter in the sub-saturation region is also important in context to core-collapse supernovae \cite{Liebendorfer_2005}, neutron star crust and giant astrophysical explosions where nuclear matter minimizes its energy by forming clusters at temperature $\approx 4$ MeV \cite{HOROWITZ200655}.

The $\gamma$ ray emission is the dominating process in the nucleus at low excitation energy, where nuclear levels are well resolved. As excitation energy increases slightly, the nuclear energy levels are substantially modified. The single-particle energy states become degenerate and nuclear shells start melting leading to a spherical nucleus after a  temperature usually known as shell melting temperature T$_m \approx 1-2$ MeV \cite{Quddus_2018}. Further increase in temperature leads to nucleon emission, which is generally studied within the framework of nuclear statistical equilibrium. On further heating, the nucleon evaporation turns violent, and at a certain limiting temperature, T$_{l}$, a new decay channel known as multi-fragmentation becomes dominant. This T$_{l}$ was found to be $\approx 5.6$ MeV for the mass region A $\approx$ 90 in ALADIN experiment \cite{PhysRevLett.102.152701}.  Nuclear multi-fragmentation occurs in the region of spinodal or phase instability boundary in the nuclear matter phase diagram \cite{vishalsymmetric}. The nucleus, which resembles a hot liquid drop, expands because of thermal pressure and moves to the spinodal region where it is surrounded by a nucleon gas.  As the spinodal is the region of instability, the nucleus explodes violently and the process is known as multi-fragmentation at freeze-out volume $\approx 7V_0$ \cite{karnau2003} 

There have been several qualitative attempts to study the limiting temperature of nuclei in terms of Coulomb instability, where the  EoS of infinite matter is taken from  various  theoretical frameworks such as Skyrme effective NN interaction \cite{PhysRevC.44.2505, PhysRevC.69.014602}, microscopic EoS such as Friedman and Pandharipande, finite temperature relativistic Dirac-Brueckner, chiral perturbation theory \cite{microscopic, m1, NICOTRA2005118}, EoS considering the degeneracy of the Fermi system \cite{PhysRevC.39.169} relativistic calculations using quantum hadrodynamics and Thomas-Fermi approach \cite{PhysRevC.47.2001, PhysRevC.49.3228, PhysRevC.55.R1641},  Gogny interactions \cite{PhysRevC.54.1137},  chiral symmetry model \cite{PhysRevC.59.3292}.  Some calculations  have been carried out by analyzing the plateau in caloric curve obtained from various experimental observations \cite{PhysRevC.65.034618}. These calculations  give a qualitative picture of $T_{l}$ and it is seen that $T_{l}$ is  model dependent and hence needs to be investigated for appropriate outcome. 

To understand the properties of LGPT in nuclei and most importantly the temperature at which the nucleus undergoes multi-fragmentation and loses its entity, we use one of the most successful effective relativistic mean-field (E-RMF) theory \cite{g2}. The E-RMF is the effective theory of hadrons as per quantum chromodynamics (QCD), which successfully explains the nuclear matter properties from finite nuclei to the neutron star and gives valuable inputs in the supernovae simulations. The E-RMF formulation calculates the volume energy of infinite nuclear matter on which the finite size corrections: surface, symmetry, Coulomb  are added to evaluate the properties of a realistic nucleus.  The idea behind using the E-RMF framework for the bulk volume energy part is that the nuclear drop is usually surrounded by a nucleon gas in complete thermodynamic equilibrium. To calculate the properties of such a system, one usually needs to solve the Gibbs conditions \cite{vishalasymmetric} where it is expected that the same equation of state (EoS) are used for  the gaseous as well as the liquid phase. 

The aim of present study is twofold: First, we investigate the properties of hot isolated nuclear drop by studying the variation of thermodynamic variables such as excitation energy, entropy, level density, fissility  etc. We compare them with available experimental or microscopic theoretical calculations \cite{BONCHE1984278, Quddus_2018}.  The second and  important part of this work is the qualitative analysis of the limiting temperature of a hot nucleus.
In HIRs, nuclei can be heated to their limiting temperature which provides an opportunity to investigate the collective motion of nucleons, and their highly chaotic and disordered behavior at high excitation energy. We use E-RMF parameter sets namely FSUGarnet, G3, IOPB-I, and most successful NL3 \cite{iopb} for the volume energy of a  nucleus. The temperature-dependent surface energy term depends on the  $T_c$ which is calculated for these individual E-RMF parameter sets. In the analysis of critical properties of infinite nuclear matter using these E-RMF sets in \cite{vishalsymmetric}, we found that the  $T_c$ is not a well-constrained quantity and the majority of E-RMF sets that satisfy the relevant observational and experimental constraints on EoS underestimates it.  Since the experimental value of  $T_c$ is calculated by extrapolating the data from multi-fragmentation reaction data on finite nuclei, it is interesting to see the variation of $T_{l}$ of finite nuclei using different E-RMF forces.  To further generalize the relationship between various saturation properties of infinite nuclear matter, its critical properties, and the limiting properties of a hot nucleus, we have used fifteen parameter sets that lie within the allowed incompressibility range and satisfy other constraints  \cite{duttra}. An effort is made to establish correlations among these properties.

The paper is organized as follows: In Section \ref{formalism}, we have discussed the formalism to calculate the energy of a finite nucleus from infinite nuclear matter . In subsequent subsections \ref{eenergy},  and  \ref{limtformalism} we discuss the formalism for the excitation energy, fissility parameter, and limiting temperature along with the lifetime of the hot nuclear liquid drop. In section \ref{results}, we have discussed results related to  various properties of hot nucleus. Finally, we summarise our results in Section \ref{conclusion}.

\section{\label{formalism} Theoretical formalism}

\subsection{\label{freeenergyformalism} From infinite matter to finite nuclei}
We consider a nucleus to be a liquid drop and resort to the conventional liquid-drop model to  define the free energy of the drop with given mass number A, proton number Z, and neutron number N as 

\begin{eqnarray}
\label{freeenergy}
F_A(\rho,T)=&\mathcal{F}_v(\rho,T)A+\mathcal{F}_{corr}(\rho,T),
\end{eqnarray}

Where $\mathcal{F}_v(\rho, T)$ is the free energy of infinite symmetric nuclear (SNM) matter calculated within the effective-relativistic mean-field theory (E-RMF) corresponding to the volume and $\mathcal{F}_{corr}$ is the finite size correction due to surface, symmetry, and Coulomb effects and is written as

\begin{eqnarray}
\begin{aligned}
\label{freecorr}
\mathcal{F}_{corr}(\rho,T)=&f_{surf}(\rho,T)4 \pi R^2+f_{sym}(\rho,T)\frac{(N-Z)^2}{A}\\
&+f_{Col}.
\end{aligned}
\end{eqnarray}

Here $R$ is the radius of the drop and is defined as

\begin{equation}
    R=\qty(\frac{3 A}{4 \pi \rho(T)})^{1/3}.
\end{equation}

The coefficient of free surface energy (FSE) ($f_{surf}(\rho, T)$) is a crucial parameter that introduces the surface and is assumed to be factorized and density-dependent \cite{RAVENHALL1983571}. This is written as 

\begin{eqnarray}
\begin{aligned}
\label{surffreeenergy}
f_{surf}(\rho,T)=\alpha_{surf}(\rho_0,T=0)\mathcal{D}(\rho)\mathcal{Y}(T).
\end{aligned}
\end{eqnarray}

Here, $\alpha_s(\rho_0,T)$ is the surface energy coefficient at saturation density ($\rho_0$) of infinite SNM. As the density of liquid evolves, the surface energy should change. Therefore the density dependence is taken from and is written as \cite{BLAIZOT1980171}
\begin{equation}
    \mathcal{D}(\rho)=1-\frac{\mathcal{K_\rho}}{2}\qty(\frac{\rho-\rho_0}{\rho_0})^2
\end{equation}

The temperature dependence of the coefficient of FSE is another significant parameter that ensures that the surface tension vanishes above a certain temperature $T_c$. In this work, we use two parametrizations of the temperature dependence of surface energy which are widely used in the calculation of multi-fragmentation in nuclei and structure of neutron star crust.  The first expression is taken from \cite{ravenhall} which takes into account the plane sharp interface between liquid and gaseous phase of nuclear matter in equilibrium. It is written as

\begin{eqnarray}
\label{s1}
\mathcal{Y}(T)&=&\qty(\frac{T_c^2-T^2}{T_c^2+T^2})^\frac{5}{4}.
\end{eqnarray}

The second expression is derived based from the semiclassical modified Seyler-Blanchard interaction and takes the form \cite{seyler} as
\begin{eqnarray}
\label{s2}
\mathcal{Y}(T)&=&\qty(1+1.5\frac{T}{T_c})\qty(1-\frac{T}{T_c})^{\frac{3}{2}}.
\end{eqnarray}

In these expressions, $T_c$ is the critical temperature of liquid-gas phase transition in infinite SNM. $\alpha_s(\rho_0,T)$ is taken as 1.15 MeV fm$^{-2}$ and $\mathcal{K_\rho}$ is a dimensionless constant taken to be 5.0 as prescribed in \cite{jnde}. 

The  coefficient of free symmetry energy (FSYE) ($f_{sym}(\rho, T)$) which depend on the mass number of liquid drop is written as

\begin{equation}
    f_{sym}(\rho,T)=\alpha_{sym}(\rho,T=0)\mathcal{G}(T)\qty(\frac{\rho}{\rho_0})^\Gamma.
\end{equation}

Here, $\alpha_{sym}(\rho,T=0)$ is further defined as

\begin{equation}
    \alpha_{sym}(\rho,T=0)=\frac{J}{1+ \mathcal{C}A^{-1/3}},
\end{equation}
where $J$ is the symmetry energy of cold SNM and is taken as $31$ MeV and $\mathcal{C}=2.4$. The dependence of $f_{sym}(\rho, T)$ on the temperature is ensured using the function $\mathcal{G}(T)$ in line with the infinite matter calculations that suggest that free FSYE increases with temperature \cite{vishalasymmetric}. It  is taken in a schematic form as \cite{gt}

\begin{equation}
    \mathcal{G}(T)=(1+\mathcal{X}_1T+\mathcal{X}_2T^2+\mathcal{X}_4T^4),
\end{equation}

where $\mathcal{X}_1=-0.00848$, $\mathcal{X}_2=0.00201$, $\mathcal{X}_4=0.0000147$ with dimensions as relevant power of unit of temperature. The density dependence is ensured with the $\Gamma=0.69$ in congruence with the experimental observations \cite{gamma}. The free Coulomb energy FCE which is otherwise absent in the infinite matter is responsible for the Coulomb instability of the liquid drop. It is taken as \cite{SAUER1976221}

\begin{equation}
    f_{Col}=\frac{3}{5}\frac{Z^2e^2}{R}\qty(1-\frac{5}{2}\qty(\frac{b}{R})^2),
\end{equation}

where b is the surface thickness which is also a temperature-dependent quantity taken as

\begin{equation}
    b \approx 0.72(1+0.009T^2).
\end{equation}

The ratio $\frac{b}{R}$ increases with temperature resulting in the reduction of Coulomb free energy in addition to that arising from the expansion of bulk matter. We do not include the exchange term in Coulomb free energy due to its low contribution. In the construction of the liquid drop, we do not include other finite-size effects such as pairing and shell corrections because they become insignificant for temperature $>1-2$ MeV due to shell melting.

\subsection{\label{ermf} E-RMF at zero and finite temperature  }
 The relativistic mean-field model  (RMF) treats nucleons as Dirac particles that interact in the relativistic covariant way by exchanging virtual mesons namely, isoscalar-scalar $\sigma$ meson, isoscalar-vector $\omega$ meson,  isovector-vector $\rho$ meson, and isovector-vector $\delta$ mesons.  
Further modification in the RMF model leads to effective relativistic mean-field formalism (E-RMF) which has the advantage that one can ignore the renormalization and divergence of the system. In E-RMF, the Lagrangian contains an infinite number of terms consistent with the underlying QCD Symmetries. The ratio of meson fields to the nucleon mass is used for the expansion and truncation scheme.  Taking recourse to the naturalness and naive dimensional analysis (NDA), it is possible to truncate the Lagrangian at the given level of accuracy. The detailed formalism and theoretical background of E-RMF can be found in \cite{vishalasymmetric, iopb, Quddus_2018, vishalsymmetric, BISWAL2020122042, bka}  and here we present a general outline of the formalism. The typical E-RMF Lagrangian fro infinite nuclear matter is written as

\begin{equation}
\begin{aligned}
\label{rmftlagrangian}
&\mathcal{E}=\psi^{\dagger}(i\alpha.\grad+\beta[M-\Phi(r)-\tau_3D(r)]+W(r)+\frac{1}{2}\tau_3R(r)\\&
+\frac{1+\tau_3}{2} A(r))\psi + \qty(\frac{1}{2}+\frac{k_3\Phi(r)}{3! M}+\frac{k_4}{4!}\frac{\Phi^2(r)}{M^2})\frac{m^2_s}{g^2_s}\Phi(r)^2\\&
-\frac{\zeta_0}{4!}\frac{1}{g^2_\omega}W(r)^4\-\frac{1}{2}\qty\Big(1+\eta_1\frac{\Phi(r)}{M}+\frac{\eta_2}{2}\frac{\Phi^2(r)}{M^2})
\frac{m^2_\omega}{g^2_\omega}W^2(r)\\&
-\frac{1}{2}\qty\Big(1+\eta_\rho\frac{\Phi(r)}{M})\frac{m^2_\rho}{g^2_\rho}R^2(r)-\Lambda_\omega(R^2(r)W^2(r))\\&
+\frac{1}{2}\frac{m^2_\delta}{g^2_\delta}(D(r))^2.
\end{aligned}
\end{equation}

Here $\Phi(r)$, W(r), R(r), D(r) and A(r) are the fields corresponding to $\sigma$, $\omega$, $\rho$ and 
$\delta $ mesons and photon respectively. The $g_s$, $g_{\omega}$, $g_{\rho}$, $g_{\delta}$ and $\frac{e^2}{4\pi }$ 
are the corresponding coupling constants and $m_s$, $m_{\omega}$, $m_{\rho}$ and $m_{\delta}$ are the 
corresponding masses. 
The  zeroth  and the third component  of energy-momentum tensor yields the energy and pressure density \cite{iopb, vishalsymmetric}.
For cold matter i.e. T=0 case, the complete field equations and related density, energy and pressure integrals are well given in \cite{iopb,g3} . At $T \ne 0$, the energy and pressure for finite temperature can be written by using the concept of canonical thermodynamic potential $\Omega$ which are also documented in \cite{vishalsymmetric, vishalasymmetric}. The Dirac effective mass which is calculated self consistently is written as 

\begin{eqnarray}
\label{effmass}
&M^*_{n/p}=M-\Phi(r) \pm  D(r).
\end{eqnarray}

\subsection{\label{eenergy} Excitation energy, level density and  fissility parameter}

The binding energy $E(T)$ of a liquid-drop with given A and Z can be found by minimizing Eq. \ref{freeenergy} to obtain the density of a nucleus at a given temperature. The excitation energy then attain a simple form as $E^*(T)=E(T)-E(T=0)$, which essentially signifies the difference of binding energy of ground level to that at any given temperature.  Here the energy can be determined from the relation
\begin{equation}
    E(T)=\mathcal{F}(T)+TS.
\end{equation}

The inter-relationship between temperature, excitation energy, and entropy which determine the level density parameter (a) is written as \cite{baym2008landau}

\begin{equation}
\label{lebeldensityeq}
E^*=aT^2, \hspace{0.5cm} S=2aT, \hspace{0.5cm} S^2=4aE^*.
\end{equation}

In a heavy nucleus, The competition between Coulomb and surface energy determines the fissility of the nucleus: the larger the ratio, the smaller is the fission barrier.
The fissility parameter  is given by dimensionless parameter $x(T)$ which is defined as \cite{SAUER1976221}

\begin{equation}
\label{fissilityeq}
    x(T)=\frac{\mathcal{F}^0_{Col}}{2\mathcal{F}^0_{s}},
\end{equation}

here superscript signifies the spherical drop. We then define the fission barrier or potential energy of deformed drop in terms of standard liquid-drop conventions as
 
\begin{equation}
\label{fbeq}
    \mathcal{B}_f(T)=((B_s-1)+2x(T)(B_c-1)).
\end{equation} 
Here, $B_s$ and $B_c$ are the surface and Coulomb energy at saddle point in the units of surface and Coulomb free energy respectively. Values of $B_s$ and $B_c$ can be determined from \cite{nix} where these values are tabulated against fissility parameter $x(T)$.

\subsection{\label{limtformalism} Limiting temperature}

The most important aspect of the thermodynamics of a finite nucleus is its multi fragmentation which can be explained in terms of liquid-gas phase transition. 
We consider the nucleus to be a spherical drop of liquid surrounded by a gas of nucleons under the assumptions  that the hot nucleus at a temperature T is surrounded by homogeneous gas of symmetric nuclear matter in a complete mechanical and chemical thermodynamic equilibrium with no exchange of particle.  A sharply defined surface separates the liquid and gaseous phase and there is no interaction between nucleons in the gaseous and liquid phase so that the gas remains unchanged and without Coulomb effect. These approximations then lead us to the following modified phase equilibrium condition similar to the infinite matter case.

\begin{subequations}

\label{coexcondition}
\begin{eqnarray}
&P_0^g(\rho^g,T)=P^l_0(\rho^l,T)+\delta P^l ,\\
&\mu_{p0}^g(\rho^g,T)=\mu_{p0}^l(\rho^l,T)+\delta \mu_p^l.
\end{eqnarray}
\end{subequations}

Here, 0 in the subscript refers to the bulk matter conditions, and $\delta P^l$ and $\delta \mu_p^l$ are the pressure and chemical potential corrections which are given as \cite{BANDYOPADHYAY19901}

\begin{subequations}

\label{correction}
\begin{eqnarray}
&\delta P^l=-\rho^2\big(\frac{\partial \mathcal{F}_{corr}}{\partial \rho}\big)\big|_{T,N,Z},\\
&\delta \mu_p^l=\big(\frac{\partial \mathcal{F}_{corr}}{\partial Z}\big)\big|_{T,N,\rho}.
\end{eqnarray}
\end{subequations}
 Where, $\mathcal{F}_{corr}$ is defined in Eq \ref{freeenergy}. The expressions for other thermodynamical quantities such as critical temperature (T$_l$), flash temperature (T$_f$), etc can be found in \cite{vishalasymmetric, vishalsymmetric} and they are used similarly in this work. The external nucleon gas also defines the stability of a hot nuclear liquid drop. In this context, we define the lifetime of a hot drop by using the concept of statistical average and assuming neutron emission to be the dominant process along with neglecting the energy dependency of the cross-section as \cite{BONCHE1984278} as

\begin{equation}
\label{timeeq}
    \frac{1}{\tau}=4 \pi \gamma \frac{1}{h^3} 2 m (kT)^2 \sigma \exp{\frac{\mu_n}{kT}},
\end{equation}
where $\gamma$ is the spin degeneracy and $\sigma$ is taken to be geometric  cross section.

\section{\label{results} Results and Discussion}

In this section, we present the results of our calculation of a hot nucleus.  We use FSUGarnet, G3, IOPB-I, and  NL3 E-RMF \cite{iopb} parameter set . These E-RMF forces are known to reproduce the properties of finite nuclei as well as infinite nuclear matter \cite{vishalsymmetric, vishalasymmetric, Quddus_2018, BISWAL2020122042, iopb}. They also satisfy the relevant constraint on EoS such as incompressibility, symmetry energy, slope parameter, etc., and observational constraints like Flow and Kaon experiments \cite{Quddus_2018}. In \cite{vishalsymmetric, vishalasymmetric} we have discussed in detail the critical properties of SNM using these parameters and,  here we extend those to the finite nuclei case.  In our calculations, for a fixed nuclear system and E-RMF parameter set, we use two parametrization  i.e. Eq. \ref{s1} and Eq. \ref{s2} for surface energy. These parameterizations are widely used in the astrophysical \cite{PhysRevC.79.035804} and statistical calculations \cite{LI201618} and are used here for comparison . We compare the results in reference to the  properties of nuclei at finite temperature and consequently study the role of critical temperature of infinite matter. This section is divided into three subsection. We discuss the caloric curve and related aspects in section \ref{excitation} and limiting temperature in section \ref{limitingtemperature}. In section \ref{correlation} we establish the correlation among various zero and finite temperature properties. 

\subsection{\label{excitation} Excitation energy, level density and fissility}

\begin{figure}
    \centering
    \includegraphics[scale=0.35]{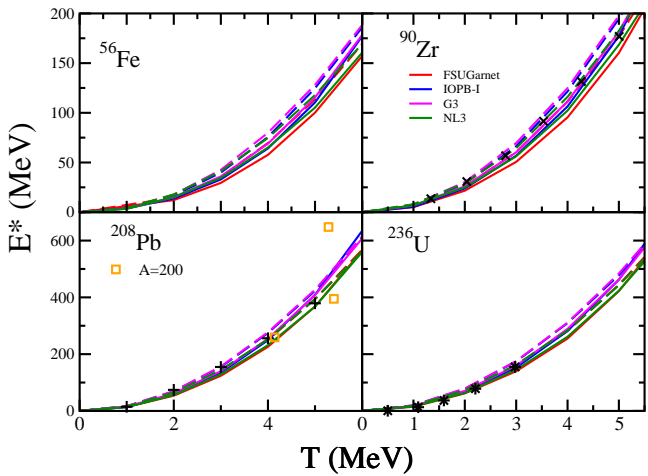}
    \caption{The excitation energy of $^{56}$Fe, $^{90}$Zr, $^{208}$Pb and $^{236}$U as a function of temperature for FSUGarnet, IOPB-I, G3 and NL3 sets. The solid lines represents calculation from Eq. \ref{s1} and dashed lines are from Eq. \ref{s2}. The theoretical data in black cross is taken from \cite{carlson2016}, plus \cite{BONCHE1984278} from  and star from \cite{Quddus_2018}. The experimental values for A $\approx$ 200 are taken from \cite{a200}.}
    \label{ex}
\end{figure}

We begin with the discussion of the caloric curve which is the relation between excitation energy and temperature for the three isolated spherical nuclei i.e. $^{56}$Fe, $^{90}$Zr, $^{208}$ Pb and  $^{236}$U which is formed when thermally fissile $^{235}$U absorb a thermal neutron. In experiments, the temperature of the nucleus is not measured directly and it is calculated using  excitation energy which can be obtained  using resonance or energy of evaporation residue. Above mentioned   nuclei are most  studied nuclear systems and their microscopic calculations are available in literature.  
Fig \ref{ex}  shows the caloric curve for these nuclei using the four E-RMF sets FSUGarnet, IOPB-I, G3, and NL3.

The estimations of theoretical caloric curves from the E-RMF are in reasonable agreement with microscopic calculations \cite{carlson2016, BONCHE1984278, Quddus_2018}. The experimental value for mass A $\approx$ 200 extracted from \cite{a200} also align with our calculations for T $<5$ MeV. The deviation at higher temperature and excitation energy may be associated with the production of heavier particles in the multi-fragmentation process which may change the 
 energy of the system. The behavior of different parameter sets is tightly constrained and the spread of curves becomes narrower as one moves from $^{56}$Fe to $^{208}$Pb. The effect of different parametrization of surface energy from Eq.\ref{s1} and \ref{s2} is also visible. Eq. \ref{s2} derived from the semi-classical Seyler-Blanchard interaction estimate a steeper slope for caloric curve as compared to the Eq \ref{s1} based on thermodynamic equilibrium of sharp interface between liquid and gaseous phase. It is because the Eq. \ref{s1} estimates relatively lower surface energy at any given temperature.

For a particular nucleus, the G3 set with the largest effective mass ($m^*/m$=0.699) estimates the steepest caloric curve while the FSUGarnet with the smallest ($m^*/m$=0.578) corresponds to the softest caloric curve. The effective mass in E-RMF  formalism is determined from the strength of scalar field because of  NN interaction. The G3 set due to small scalar self couplings $k_3, k_4$ and  scalar-vector cross couplings $\eta_1 , \eta_2$ estimate the softest scalar field while the FSUGarnet yields the stiffest scalar field. The scalar field consequently determine the mechanical properties of the system and therefore, the effective mass becomes a crucial saturation property at finite temperature. The effective mass which is obtained self consistently also determine the chemical potential and kinetic energy of nucleons which are essential input for the thermal properties calculations. Furthermore the G3 set  estimate the softest repulsive contribution arising from the vector self coupling $\zeta_0$. The combine effect of scalar and vector field determine the critical temperature.  The parameter set G3 and FSUGarnet  estimate  the largest and smallest $T_c$ among these four sets (see table \ref{criticalparamaters}). Therefore, in finite nuclei, the thermal contribution of energy essentially depends on the combined effect of effective mass, $T_c$ and the zero-temperature EoS. It may be noted  that the saturation properties are not unique and different combination of mesons coupling can yield the similar nuclear matter properties. Therefore, it is  relevant to analyse the finite temperature properties of the nuclear matter in terms of saturation properties and not the coupling constants.

\begin{figure}[h]
    \centering
    \includegraphics[scale=0.35]{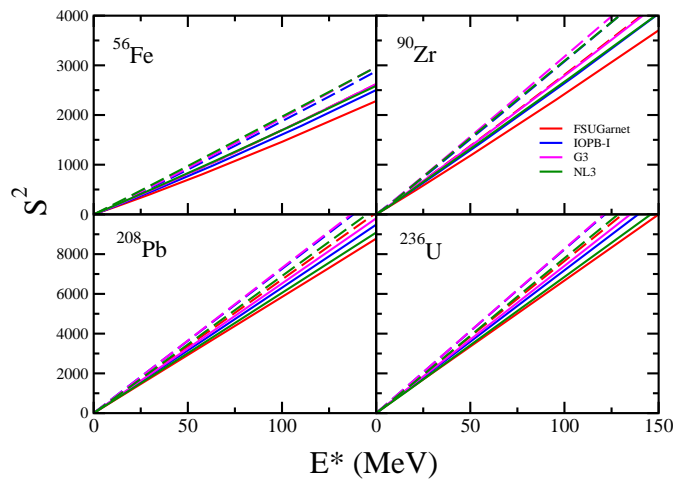}
    \caption{The Relation between square of entropy and excitation energy for the systems as in Fig \ref{ex}. }
    \label{entropy}
\end{figure}

In the Fermi gas model, the point of minimum entropy in the transition state nucleus corresponds to its minimum excitation energy (E$^*$) \cite{PhysRevLett.25.386}. Therefore we show the relation of the square of entropy and E$^*$ in Figure \ref{entropy} for the systems considered in Fig. \ref{ex}. The square of entropy increases monotonically with  the E$^*$ signifying a disordered and chaotic nucleus. The disorder increases with mass number implying a more violent multi-fragmentation process once the nucleus reaches its limiting temperature T$_l$.   
Eq. \ref{s2} estimates higher entropy at a given E$^*$ as compared to  \ref{s1}. For a particular nucleus, the spread of different E-RMF sets increases with E$^*$. This effect can  be  attributed to the effective mass and $T_c$ of a particular E-RMF parameter. In our model, we have not considered the shell correction which deviates the straight-line behaviour of this curve at low temperature, where shell structure is still intact. These shells melt at around $E^*\approx40$ MeV or $T\approx1-2$MeV \cite{PhysRevC.91.044620}. After this temperature, the nucleus is highly disordered where nucleons are constantly trying to push out from the nuclear boundary which is ensured by the surface as in Eq. \ref{surffreeenergy}. The behaviour of $S^2$ is in agreement with results in  \cite{carlson2016, Quddus_2018}. 
 
\begin{table}
    \centering
        \caption{The level density parameters obtained using different expression of Eq.  \ref{lebeldensityeq} for the FSUGarnet, IOPB-I, G3 and NL3 parameter set.}
    \begin{tabular*}{\linewidth}{c @{\extracolsep{\fill}}  cccccc}

    \toprule
 Element             & Forces    &  \multicolumn{3}{l}{ $a$ (MeV$^{-1}$) Using Eq. \ref{s2}} \\ \midrule
                    &           & ${E^*}/{T^2}$    & $S^2/4E^*$   & $S/2T$ \\ 
\midrule                    
\multirow{4}{*}{$^{56}$Fe} & NL3       & 4.695 & 4.931  & 4.789\\
                    & FSUGarnet & 4.582 & 4.323 & 4.357  \\
                    & IOPB-I    & 5.033 & 4.789 & 4.808  \\
                    & G3        & 5.149 & 4.942  & 4.963 \\ \midrule
\multirow{4}{*}{$^{90}$Zn} & NL3       & 7.267  & 7.740 & 7.491 \\
                    & FSUGarnet & 7.102 & 7.185 & 7.072  \\
                    & IOPB-I    & 7.8123  & 7.857 & 7.758  \\
                    & G3        & 7.872 & 8.065 & 7.930 \\
\midrule                    
\multirow{4}{*}{$^{208}$Pb} & NL3       & 15.683 & 17.030 & 16.394  \\
                    & FSUGarnet & 15.752  & 16.725  & 16.233\\
                    & IOPB-I    & 16.998  & 18.040 & 17.531\\
                    & G3        & 17.126 & 18.191 & 17.683\\
\midrule                    
\multirow{4}{*}{$^{236}$U}  & NL3       & 17.64   & 19.19   & 18.4632 \\
                    & FSUGarnet & 17.761 & 18.946 & 18.353\\
                    & IOPB-I    & 19.196 & 20.400 & 19.818 \\
                    & G3        & 19.296 & 20.532 & 19.949 \\
\bottomrule
    \end{tabular*}
\label{avalue}
\end{table}

The caloric curve gives us the opportunity to study the level density parameter (a) which plays a crucial role to understand the particle spectra and nuclear fission. Level density signifies the available excited state level at a given energy. In order to study the level density we use Eq. \ref{lebeldensityeq} and fit them for the value of $a$ with R-squared value $>$ 0.99. The level density parameters obtained using different expression of Eq. \ref{lebeldensityeq} are listed in  Table \ref{avalue}. The level density calculated from all the three equations in Eqs. \ref{lebeldensityeq} are comparable. A larger effective mass and T$_c$ corresponds to the larger level density as in the case of G3.  These calculations are performed using   Eq. \ref{s1}. On the other hand Eq. \ref{s2} estimates lower magnitude of level density although the trend remain same. The value of level density lie  within the  empirical relations  $A/11.93$ from \cite{11.93} and A/14.75 from \cite{14.75}. Nuclear level density can also be studied in terms of temperature where one can take the relevant ratio in a straightforward manner for eg. $a=E^*/T^2$ at a particular temperature.  The G3 set with largest effective mass yield the largest temperature-dependent level density. The above analysis of thermal properties advocates the importance of effective mass over other saturation properties. 

Now we shell discuss the temperature dependence of fissility and fission barrier. Fissility characterizes the stability of a charged nuclear drop against fission. In general, when Coulomb free energy $\mathcal{F}_{col}$ becomes twice the surface free energy $\mathcal{F}_{surf}$, the spherical liquid drop become critical towards spheroidal deformation and split into two equal parts. This feature is excessively used in the equilibrium condition determining the structure of neutron star crust or supernovae explosion (see Eq. 44 in \cite{PhysRevC.79.035804}).  One thing to note here is that, similar to a classical liquid drop, on increasing the temperature, the nuclear liquid drop becomes more spherical \cite{Quddus_2018} i.e shell structure becomes trivial and deformations in the nucleus vanish. Therefore, a drop can not undergo spontaneous fission only by the temperature and one always needs external disturbance like a thermal neutron in case of $^{235}$U. Although, at a certain maximum temperature $T_{l}$, the nucleus will undergo multi-fragmentation process.

\begin{figure}[h]
    \centering
    \includegraphics[scale=0.35]{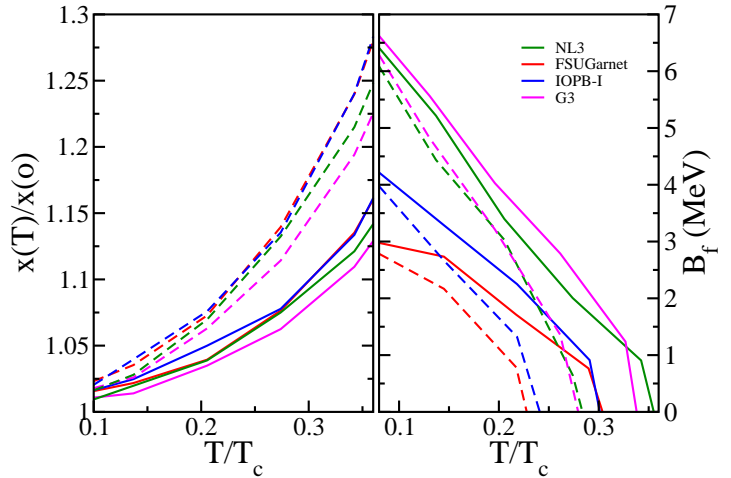}
    \caption{The fissility parameter $x(T)/x(T=0)$ as a function of $\frac{T}{T_c}$ for  $^{236}U$ using parameter sets  FSUGarnet, IOPB-I, G3 and NL3 on the left panel. Right panel show the Liquid-drop fission barrier for $^{236}U$. Solid and dashes lines have the same meaning as in Fig \ref{ex}.}
    \label{barrierfissility}
\end{figure}

We show in Fig. \ref{barrierfissility}, the variation of fissility as a function of  $T/T_c$ using Eq. \ref{fissilityeq} with different forces and both the parametrization of temperature dependence of surface energy i.e. Eq. \ref{s1} and \ref{s2}. The fissility for  $^{236}$U increases exponentially with temperature suggesting that the surface energy decreases much faster on increasing the temperature. Eq. \ref{s2} has steeper slope than Eq \ref{s1} which is again the result of lower surface energy in case of Eq. \ref{s1}. The fission barrier decreases with temperature and almost vanishes for $T/T_c$=0.4 for all the forces. G3 parameter set estimates the largest barrier and FSUGarnet the lowest which may be due to their effective mass. The effective mass controls the mechanical properties and consequently determine the equilibrium density of the nuclear liquid drop. One may notice in in fig \ref{barrierfissility} the dominant effect of $T_c$  as these quantities do not include the volume term (see Eq. 
\ref{fissilityeq}). The FSUGarnet and IOPB-I shows the similar trend with almost similar $T_c$. G3 parameter set estimates the softest fissility and largest fission barrier followed by the NL3 set as their value of $T_c$ are 15.3 and 13.75 respectively. The vanishing points of liquid-drop fission barrier are aligned with their respective value of $T_c$ (see Table \ref{criticalparamaters}).

\subsection{ \label{limitingtemperature}    Limiting temperature}

\begin{figure*}
    \centering
    \includegraphics[scale=0.6]{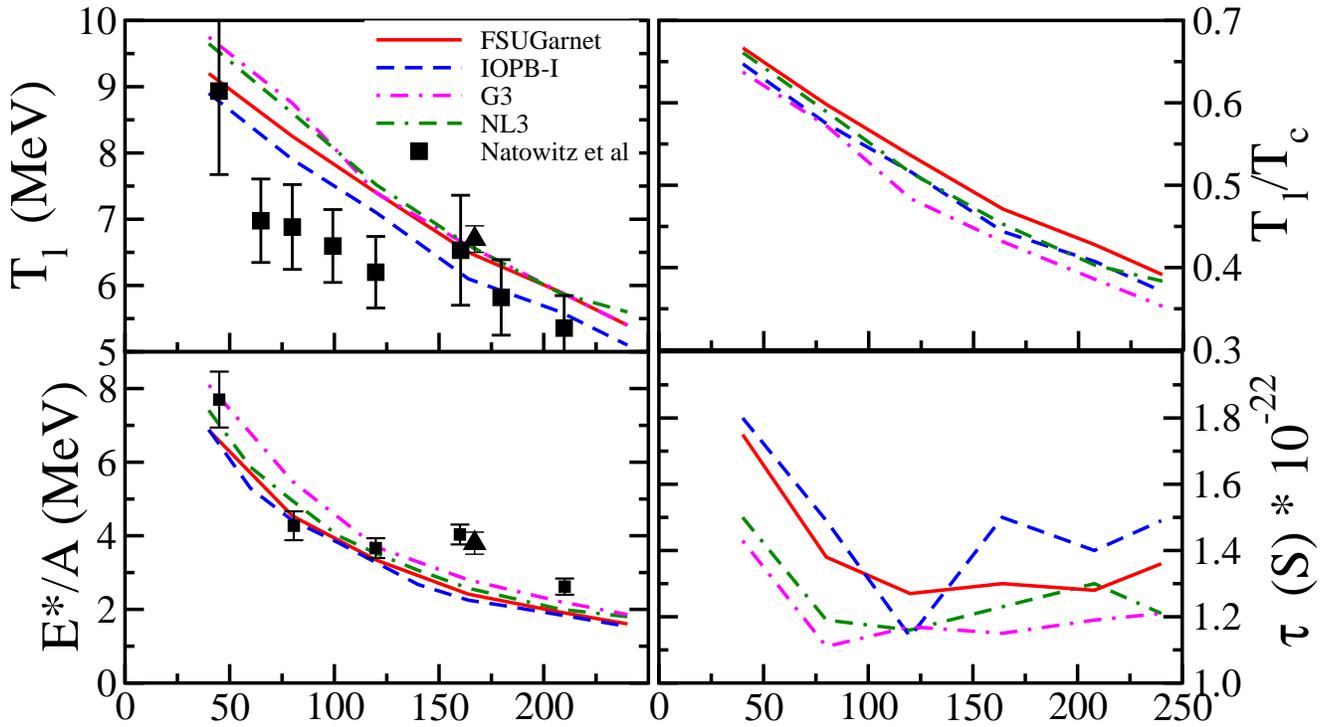}
    \caption{The limiting temperature T$_l$, the ratio of $\frac{T_l}{T_c}$, limiting excitation energy per nucleon and life time of nuclear liquid drop at the limiting temperature as a function of mass number A for the nuclei on $\beta$-stability line. The temperature dependent expression used here is Eq \ref{s1}. Experimental points in solid square are taken from \cite{natowitz} for T$_{l}$ which are calculated using double isotope yield ratio and thermal bremsstrahlung measurements and from \cite{PhysRevC.65.034618} for excitation energy. The points represented in upper triangle are taken from  the fisher droplet model derived from \cite{fisher}. }
    \label{tlim}
\end{figure*}

Determination of the temperature at which a hot nucleus drop will undergo multi-fragmentation by loosing its entity, is one of a challenging problem in nuclear physics. Experimentally it is difficult to estimate  $T_{l}$  and other related properties such as specific heat for a particular nucleus as there are large number of nucleons involved. Although, theoretically we can study these properties by applying appropriate constraints. In that context, we consider a simplistic approach to determine the $T_{l}$ of a nucleus. We employ our assumption stated in \ref{limtformalism} and solve  Eqs. \ref{coexcondition}. These Equations will not have any solution for a given T,  $\rho_v$ and $\rho_l$  for temperature greater than $T_{l}$ signifying that the nucleus can no longer exists. 

In fig. \ref{tlim} we show the variation of limiting temperature $T_{l}$, $T_{l}/T_c$, limiting excitation energy ($E^*(T_{l})/A$) and the life time ($\tau$)  of nucleus at limiting temperature as a function of mass number for the nuclei along $\beta$ stability line where the atomic number can be written as

\begin{equation}
    Z=0.5 A -0.3\cross10^{-2}A^{\frac{5}{3}}.
\end{equation}

The value of  $T_{l}$ decreases exponentially with increasing mass number as the Coulomb energy rises due to larger Z. At lower Z,  $T_{l}$ decreases at faster pace because the Coulomb component dominates the surface and symmetry energy of liquid drop. At a higher mass number, the situation becomes a little different. There is competition between Coulomb, surface, and symmetry terms. On moving from low to higher mass number along the $\beta$ stability line, the $Z/A$ ratio decreases. The decrease in the $Z/A$ ratio weakens the A dependence causing  $T_{l}$ to increases. On the other hand, the symmetry and surface energy increase with the increase in mass number which tries to bring down the $T_{l}$. For comparison we  show points determined from phenomenological analysis \cite{natowitz, PhysRevC.65.034618, fisher} for the T$_{l}$ and $E^*(T_{l})$. The results from E-RMF forces are within reasonable agreement. 

\begin{figure}
    \centering
    \includegraphics[scale=0.3]{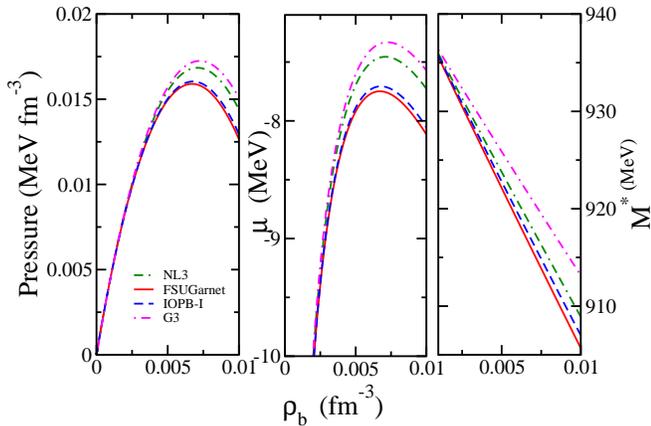}
    \caption{EoS, Chemical potential and effective mass at low density at T=5 MeV for the FSUGarnet, IOPB-I, G3 and NL3 parameter sets.}
    \label{forceprop}
\end{figure}

The value of $T_{l}$ for a particular nucleus and a particular EoS depends on the $T_{c}$ of infinite nuclear matter  and the low density ( $\rho_0<0.01$) variation of EoS which determine the properties of surrounding gaseous phase.  Since the finite-size corrections are employed externally, they are the same for every EoS. To understand the effect of EoS, we plot in Fig. \ref{forceprop}
the EoS, chemical potential ($\mu$), and effective mass ($M^*$) calculated using
the FSUGarnet, IOPB-I, G3, and NL3 parameter sets for the density range significant for nuclear vapor surrounding the hot nucleus. Chemical potential is a function of temperature-dependent effective mass which consequently determines the chemical equilibrium between nuclear gas and nuclear drop. The IOPB-I and FSUGarnet have similar ground state saturation properties and they have similar behavior at T=5 MeV.  The incompatibilities of the NL3 and G3 sets are 271.38 and 243.96 MeV respectively but their behavior is opposite in the low-density regime. G3 set estimates the maximum value of pressure, and effective mass at any given density. This is the reason G3 set have larger value of T$_c$ than the NL3 set. This trend in Fig \ref{forceprop} for different EoS,  validates the variation of   $T_{l}$ in Fig \ref{tlim}, where the magnitude of T$_l$ explicitly depends upon the low density EoS. In other word, to understand the effect of EoS on the $T_{l}$ one has to take into account the $T_{c}$ and low density behaviour of EoS instead of incompressibility at saturation.

Further the ratio $T_{l}/T_{c}$ signify the finite size effect where the limiting temperature decreases sharply as compared to the critical temperature of infinite symmetric matter. It reduces up to the 0.3$T_{c}$ for heavy nuclei. Furthermore, there is still model dependency in the $T_l/T_c$ . The larger effective mass yields smaller $T_l/T_c$ which is  clear from the fact the FSUGarnet and G3 estimate the largest and smallest $T_l/T_c$.  Limiting excitation energy per nucleon is calculated at $T_{l}$ and our calculations from E-RMF forces agree with the phenomenological calculation \cite{PhysRevC.65.034618}. We have performed these calculations using Eq. \ref{s1} as there were no significant difference between the values of $T_{l}$ calculated from Eq. \ref{s1} and \ref{s2}. However,  the Eq. \ref{s2}  estimates the  larger excitation energy for a given nucleus as compared to the Eq. \ref{s1}. The Eq. \ref{s1} and \ref{s2} are frequently used in various calculations such as statistical equilibrium analysis and supernovae matter. In that context,  these equations correctly estimate the finite nucleus observables with slight difference in magnitude.  Eq. \ref{s1} has a slight edge as it is  consistent with the surface energy estimated from thermal Hartree-Fock approximation \cite{SAUER1976221}. Our calculations show better agreement with  experimental and theoretical values when using Eq. \ref{s1} as well. However, the judicious use of these can be made depending on the problem such as supernova where the thermal energy plays a very important role.

\begin{figure*}
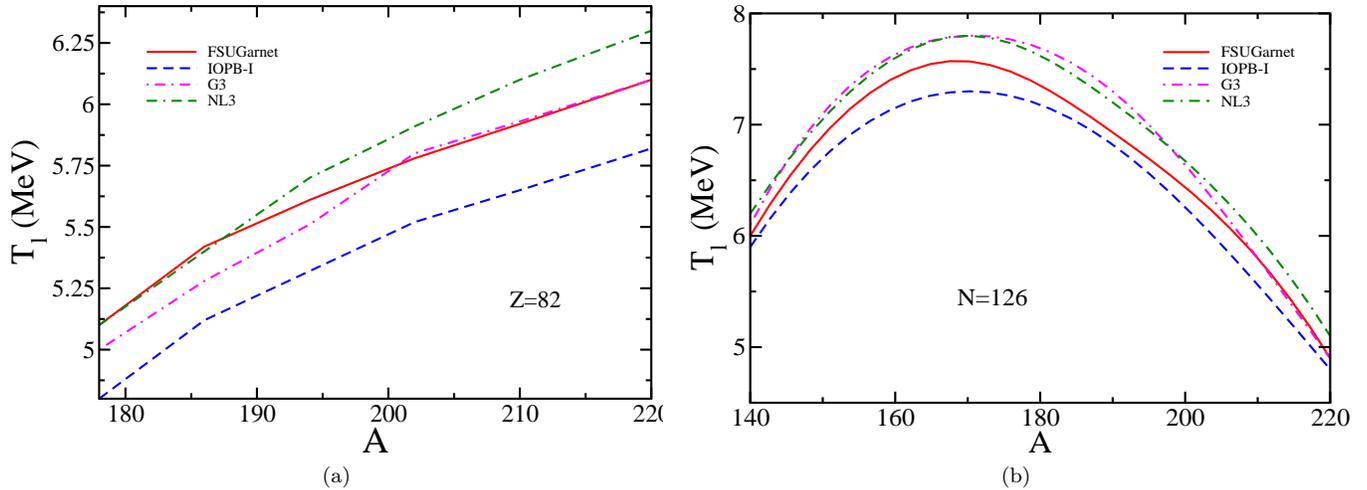

  \centering
\subfloat[]{%
  \label{fixz}
  \includegraphics[height=6cm,width=.49\linewidth]{fixz.eps}%
}\hfill
\subfloat[]{%
  \label{fixn}
  \includegraphics[height=6cm,width=.49\linewidth]{fixn.eps}%
} 
\caption{(a) Limiting temperature for fix atomic number Z=82 as a function of mass number calculated from the expression \ref{s2} (b) Same as in (a) but for fix neutron number N=126. }
\end{figure*}

To further understand the behaviour of  $T_{l}$, we calculate the lifetime of hot nucleus $\tau$  using Eq. \ref{timeeq}. As we have not considered the temperature dependence of neutron-capture cross-section, these values will slightly underestimate the lifetime but the trend will remain the same. The radius R which is the input for Eq. \ref{timeeq}   is determined after solving the coexistence Eqs. \ref{coexcondition} for a particular nucleus. We have seen that the nuclear gas surrounding the nuclear liquid plays a significant role in  determining the T$_l$. In terms of lifetime, a larger pressure and smaller density  corresponds to a less stable liquid drop and therefore, lower lifetime. The IOPB-I set that estimate the loweer T$_l$ for a given nucleus yields the higher lifetime.  We see that the lifetime $\tau$ is of the order of $10^{-22}$S at $T_{l}$ for all the nuclei on the $\beta$ stability line. Nuclei at the lower mass range are slightly more stable than heavy nuclei. This time scale is just enough for a nucleus to allow thermalization. This also states the fact that at $T_{l}$ the nucleus is highly unstable and will undergo violent multi-fragmentation which has the time scale of $10^{-22}$S \cite{KARNAUKHOV200691, lifetimevalue}.

In Fig. \ref{fixz}, the variation of $T_{l}$ is shown for a fixed atomic number Z=82 and  Fig. \ref{fixn} demonstrate the behaviour for a fixed neutron number N=126. For a fixed atomic number, the $T_{l}$ rises $\approx$ 1.5 MeV when we move from A=178 to A=220 or from  $Z/A$ = 0.46 to 0.37. The increase in $T_{l}$ with a decrease in $Z/A$ ratio is because Coulomb free energy reduces as the radius of nuclear liquid drop increases as a function of charge number. The surface energy then dominates over the Coulomb energy which helps in preserving the surface of the drop at a much higher temperature. This trend is confirmed with the non-relativistic Hartree-Fock calculation where the solution becomes unstable after a certain temperature \cite{BONCHE1985265}. When we keep the neutron number fixed, there is an interesting binodal type trend in the values of $T_{l}$ with increasing mass number. The $T_{l}$ increases with increasing Z and reaches its maximum at A $\approx$ 170. It then decreases at a faster rate on further increasing the value of Z. This effect is the result of competition between Coulomb and surface energy at lower and higher mass region. This shape of the graph then signifies that one can make nuclei in the unconventional regimes, which might not be stable at zero temperature but can exist at some higher temperature. In Figs. \ref{fixz} and \ref{fixn}, the trends of EoS are similar to the ones obtained at low density regime.

\subsection{\label{correlation}   Correlations}
In the analysis of a hot nucleus and its limiting temperature, we saw that the critical temperature $T_c$ of infinite nuclear matter affects the observables through Eq. \ref{s1} and \ref{s2}. They also depend on the properties of EoS such as effective mass and  low density behaviour of a particular EoS. The $T_c$ which is basically an inflation point on critical isotherm, is one of the most uncertain parameter in nuclear matter studies. The value of $T_c$ is an important factor in calculation of finite nuclei as well as supernovae matter and neutron star crust \cite{PhysRevC.79.035804}. Hence it becomes important to relate the $T_c$ of a particular EoS to its saturation properties. In Refs. \cite{vishalsymmetric, vishalasymmetric} we have studied  the thermodynamics of liquid gas phase transition in infinite nuclear matter using the E-RMF parameter sets used in this study. It has been observed  that the critical temperature $T_c$ is not a  well constrained quantity.  It requires a comprehensive statistical analysis of nuclear  properties at critical points and saturation properties of cold nuclear matter as their analytical relationship is difficult to establish.  For this, we take fifteen E-RMF  parameter sets  satisfying relevant constraints \cite{duttra, iopb, vishalasymmetric, Quddus_2018, bka} on EoS and first of all calculate the properties at critical point of liquid-gas phase transition in infinite matter. 

\begin{table*}[t]
    \centering
        \caption{The zero temperature incompressibility K, binding energy $e_0$, saturation density $\rho_0$, effective mass M$^*$ and  critical temperature $T_c$, pressure $P_c$, density $\rho_c$ along with flash temperature $T_f$, density $\rho_f$, incompressibility $C_f$ and effective mass at $T_c$ for infinite symmetric nuclear matter using the several forces. }
    \begin{tabular*}{\textwidth}{c @{\extracolsep{\fill}}  ccccccccccccc}
 \toprule
  Parameter & K      & $e_0$     &  $\rho_0$    &  $m^*/m$    & $T_c$    & $P_c$    & $\rho_c$     & $T_f$     & $\rho_f$     & C$_f$ & $m^*_c/m$   \\
  & MeV     & MeV     &  fm$^{-3}$    &      & MeV    & MeV fm$^{-3}$    & fm$^{-3}$     & MeV     & fm$^{-3}$     &   & \\ 
  
  \midrule
  
 G2   \cite{g2}        & 215.00 & -16.10 & 0.153 & 0.664 & 14.30 & 0.181 & 0.0432 & 11.80 & 0.080 & 0.293 & 0.879 \\
IOPB-I   \cite{iopb}   & 222.65 & -16.10 & 0.149 & 0.593 & 13.75 & 0.167 & 0.0424 & 11.20 & 0.071 & 0.286 & 0.864 \\
Big Apple \cite{bigapple}  & 227.00 & -16.34 & 0.155 & 0.608 & 14.20 & 0.186 & 0.0441 & 11.45 & 0.073 & 0.297 & 0.876 \\
BKA22   \cite{bka}   & 227.00 & -16.10 & 0.148 & 0.610 & 13.90 & 0.178 & 0.0442 & 11.33 & 0.072 & 0.290 & 0.855 \\
BKA24  \cite{bka}    & 228.00 & -16.10 & 0.148 & 0.600 & 13.85 & 0.177 & 0.0450 & 11.31 & 0.073 & 0.284 & 0.845 \\
FSUGarnet  \cite{iopb} & 229.50 & -16.23 & 0.153 & 0.578 & 13.80 & 0.171 & 0.0430 & 11.30 & 0.071 & 0.288 & 0.850 \\
FSUGold  \cite{fsugold} & 230.00 & -16.28 & 0.148 & 0.600 & 14.80 & 0.205 & 0.0460 & 11.90 & 0.074 & 0.301 & 0.844 \\
IUFSU   \cite{iufsu}   & 231.31 & -16.40 & 0.155 & 0.610 & 14.49 & 0.196 & 0.0457 & 11.73 & 0.074 & 0.296 & 0.862 \\
FSUGold2 \cite{fsugold2}  & 238.00 & -16.28 & 0.151 & 0.593 & 14.20 & 0.187 & 0.0450 & 11.51 & 0.073 & 0.293 & 0.855 \\
BKA20   \cite{bka}   & 240.00 & -16.10 & 0.146 & 0.640 & 15.00 & 0.209 & 0.0458 & 11.91 & 0.073 & 0.304 & 0.868 \\
G3    \cite{iopb}      & 243.96 & -16.02 & 0.148 & 0.699 & 15.30 & 0.218 & 0.0490 & 12.10 & 0.075 & 0.291 & 0.879 \\
NL3* \cite{nl3*}      & 258.27 & -16.31 & 0.150 & 0.590 & 14.60 & 0.202 & 0.0466 & 11.70 & 0.075 & 0.297 & 0.861 \\
Z27v1   \cite{z27v}    & 271.00 & -16.24 & 0.148 & 0.800 & 18.03 & 0.304 & 0.0515 & 13.70 & 0.077 & 0.327 & 0.914 \\
NL3  \cite{iopb}       & 271.38 & -16.29 & 0.148 & 0.595 & 14.60 & 0.202 & 0.0460 & 11.80 & 0.070 & 0.301 & 0.846 \\
TM1   \cite{tm1}      & 281.10 & -16.26 & 0.145 & 0.630 & 15.60 & 0.236 & 0.0486 & 12.09 & 0.076 & 0.311 & 0.862 \\
\midrule
Exp/Emp & 240  \cite{duttra}   & -16 \cite{vishalasymmetric} & 0.166 \cite{vishalasymmetric}  & 0.63  \cite{FURNSTAHL1998607}  & 17.9 \cite{tcexp} & 0.31\cite{tcexp} & 0.06 \cite{tcexp} & -  & -& 0.288 \cite{vishalsymmetric} & -\\
 & $\pm$ 20 & $\pm$ 1  & $\pm$ 0.019  & $\pm$ 0.05  & $\pm$ 0.40  & $\pm$ 0.07 &  $\pm$ 0.01 & -  & - & -& -\\
\bottomrule
    \end{tabular*}
\label{criticalparamaters}
\end{table*}

\begin{figure}
    \centering
    \includegraphics[scale=0.35]{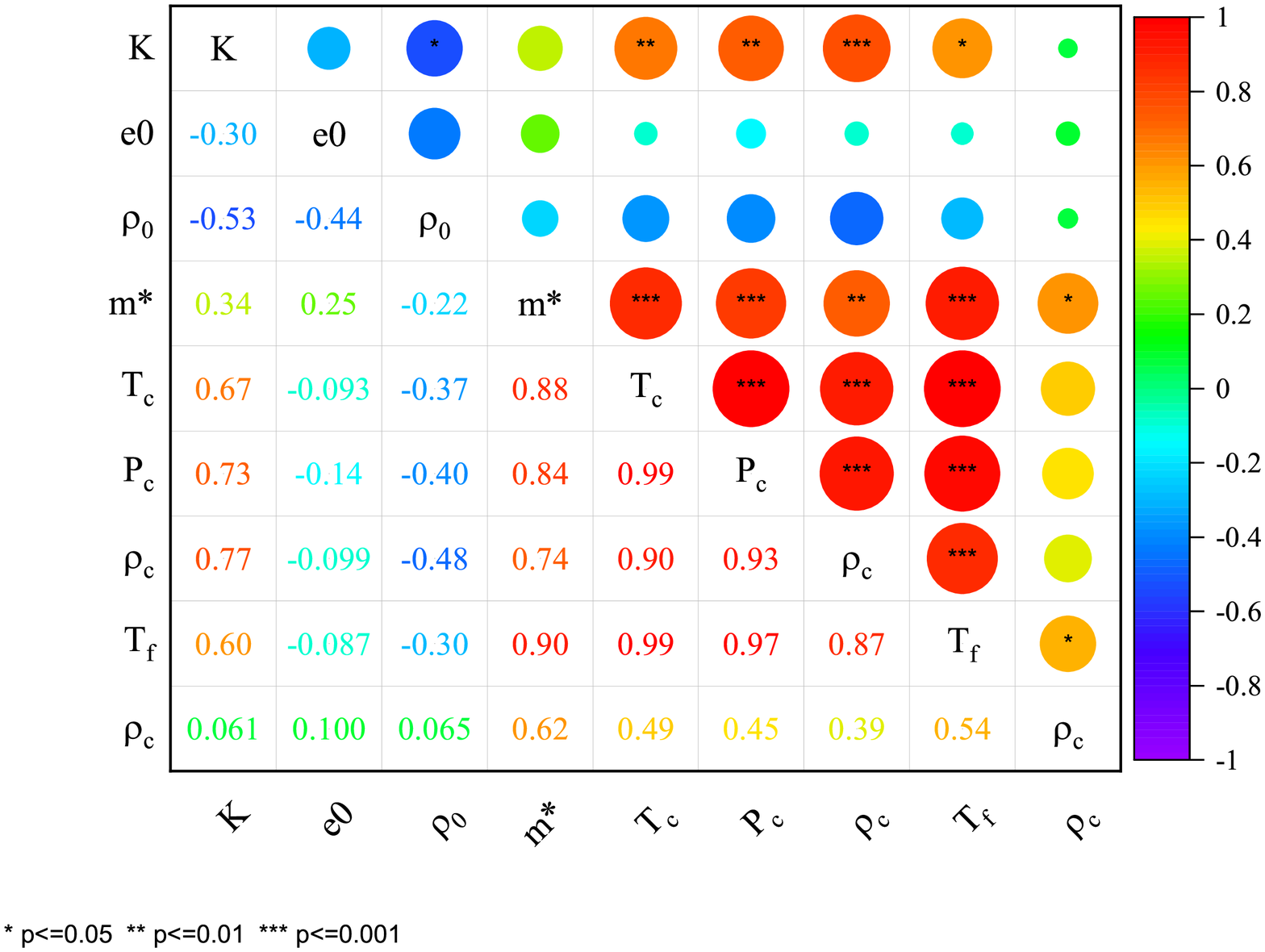}
    \caption{The Pearson correlation matrix for the critical parameter for infinite symmetric nuclear matter and some cold nuclear matter properties. The number of stars in a circle represents the p-value given at the bottom. The strength of correlation is colour mapped.}
    \label{tctp}
\end{figure}

In table \ref{criticalparamaters}, we present the saturation properties of cold nuclear matter i.e. incompressibility (K), binding energy ($e_0$), saturation density ($\rho_0$), effective mass (M$^*$) and  critical temperature ($T_c$), pressure ($P_c$), density ($\rho_c$) along with flash temperature ($T_f$), density ($\rho_f$), incompressibility factor ($C_f$) and effective mass at ($T_c$) for infinite symmetric nuclear matter using different force parameters. For further details on these quantities please see Ref. \cite{vishalsymmetric}. We have selected variety of forces with different meson couplings, which include up to the quartic order scalar and vector terms  in order to have a generalised analysis of E-RMF forces.  The E-RMF sets satisfying the allowed incompressibility range and other observational constraints underestimate the critical values of temperature, density and pressure when compared to experimental data \cite{tcexp}.

We then calculate the Pearson Correlation matrix \cite{pearson} for variables calculated in table \ref{criticalparamaters} and the results are  shown in Fig. \ref{tctp}. The colour coded correlation  matrix also show the statistical significance in form of p-value \cite{pearson} for different confidence interval i.e. 95\%, 99\% and 99.9\%. The binding energy ($e_0$) and saturation density ($\rho_0$) of cold infinite nuclear matter have very weak strength of correlation with the critical properties at finite temperature. This is against the natural intuition that binding energy of infinite matter should impact  the $T_c$.

The incompressibility on the other hand show positive correlation with critical properties i.e. $T_c$, $p_c$, $\rho_c$ and $T_f$. Although this correlation do not exceed the value of 0.77. Therefore, we can conclude that the saturation properties of cold nuclear matter do not significantly impact the value of critical parameter individually. The reason for this can be the fact that saturation properties are calculated at saturation density $\rho_0\approx$ 0.16 fm$^{-3}$, whereas, the nuclear matter convert from liquid to gaseous phase at $\approx$ 0.25-0.3 $\rho_0$. The behaviour of EoS in this density region is not always as per the properties at saturation, as noticed  in Fig \ref{forceprop}. One exception is the effective mass which shows a strong positive correlation with critical properties. This is in line with our analytical analysis of  infinite nuclear matter that finite temperature properties in E-RMF formalism are  governed by the effective mass. This behaviour is consistent with the non-relativistic formalisms as well, although the definition of effective mass is different in both the cases \cite{vishalasymmetric}.  

From Table \ref{criticalparamaters} we see that the parameter sets G3 and Z27v1 have relatively high effective mass and a high value of $T_c$. A high positive correlation between $m^*$ and $T_c$ in Fig \ref{tctp} suggests the same. Therefore, one way to construct a model at par with experimental findings is to exploit this property of effective mass. This fact was also considered in \cite{PhysRevC.94.045207}. However, the prescribed range of effective mass 0.58 $\le$ m$^*$/m $\le $ 0.68 in agreement with spin-orbit splitting experiments \cite{FURNSTAHL1998607} should be kept in mind. The Z27v1 set does not satisfy this constraint and it was also not considered in \cite{duttra}, from where the constraints on EoS are taken for this study. Therefore, no standard RMF and E-RMF parameter sets, that satisfy all the available constraints can reproduce the experimental value of the critical parameter for infinite nuclear matter and hence needs more analysis especially on the low-density regime of EoS. Moreover, the effective mass dependence of thermal properties will also be useful in the microscopic calculations, where the concept of $T_c$ is not explicitly used for the surface energy calculation.

\begin{figure}[h]
    \centering
    \includegraphics[scale=0.3]{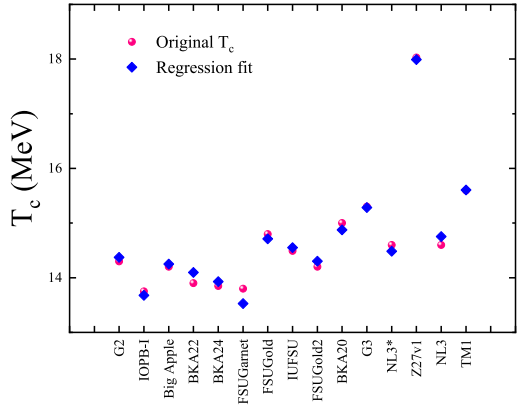}
    \caption{Actual value of $T_c$ from different forces and regression fit values calculated from Eq. \ref{regfiteq}.}
    \label{regfit}
\end{figure}

The low correlation means that the variables are acting as independent parameters. This is also justified as the properties like K, $\rho_0$, $e_0$, and $m^*$ are the inherent characteristic properties of an EoS.   The critical temperature therefore can be understood as a  result of competition between various nuclear matter observables. To demonstrate this, we construct a very simple multiple linear regression (MLR) fit of following form.

\begin{equation}
\label{regfiteq}
    T_c= \beta_0+ \beta_1K+\beta_2e_0+\beta_3\rho_0^{(1/3)}+\beta_4m^*,
\end{equation}
where, all the variables are in MeV except $\rho_0$ which is in MeV$^3$ and coefficients have relevant dimensions with $\beta_{0,1,2,3,4}$=-11.5033, 0.00201, -4.32248, -0.52433, 0.01795. These coefficients are statistically significant as well for 95 \% confidence interval.  In Fig. \ref{regfit}, we show the result of Eq. \ref{regfiteq} against the actual $T_c$ from table \ref{criticalparamaters}. The regression equation estimate the  $T_c$ excellently with R-square=0.987. The fitted regression equation suggest that the binding energy and saturation density has opposite variation with  $T_c$.  The regression equation \ref{regfiteq} is better than the empirical relations suggested in \cite{RIOS201058} based on Lattimer–Swesty and Natowitz predictions. This is because the greater degree of freedom are considered in this equation. However, this will yield a strange value of $T_c$ when all the saturation properties tend to zero. This equation gives an useful insight in the form of free coefficient $\beta_0$ which  suggests that there is a missing link between our current understanding of critical temperature and its relationship with the saturation properties. The $\beta_0$ becomes inevitable as the equation then gives a bad fitting. 

Unlike saturation and critical properties, the critical parameters are strongly correlated with each other, except the flash density $\rho_f$. The flash density seems to be model independent with standard deviation = 0.0025.  It is to note that these correlations are for the E-RMF sets considered in Table \ref{criticalparamaters} and are not universal. However, the selected parameter sets have a wide range of meson couplings and nuclear matter saturation properties. Moreover, we have presented our results with the statistical significance of Pearson  correlation to make it as general as possible. 

\begin{table}[]
    \centering
        \caption{Limiting temperature (MeV), chemical potential (MeV), pressure (MeV fm$^{-3}$), gas density ($fm^{-3}$), liquid density ($fm^{-3}$), radius ($fm$) and lifetime ($\tau\cross\exp{-22}$ Sec) of $^{208}$Pb nucleus for several forces.}
    \begin{tabular*}{\linewidth}{c @{\extracolsep{\fill}}  ccccccccc}
 \toprule
 
 Parameter  & $T_l$   & $\mu$      & P & $\rho_v$    & $\rho_l$  & R & $\tau$ \\
 
 \midrule
 
 G2            & 5.4  & -8.55  & 0.0162   & 0.0075  & 0.147 & 6.964 & 1.49 \\
IOPB-I        & 5.88 & -9.61   & 0.02     & 0.0084  & 0.143 & 7.028 & 1.29 \\
Big Apple     & 5.37 & -8.36 & 0.0191   & 0.0075 & 0.148 & 6.948& 1.47\\
BKA22         & 5.46 & -8.65  & 0.0197   & 0.0076  & 0.142 & 7.045 & 1.42\\
BKA24         & 5.51 & -8.73  & 0.0196   & 0.0075  & 0.142 & 7.045 & 1.40\\
FSUGarnet     & 5.9  & -9.48   & 0.024    & 0.0082  & 0.148 & 6.948 & 1.28 \\
FSUGold       & 5.92 & -9.21   & 0.0239   & 0.0085  & 0.143 & 7.028 & 1.18\\
IUFSU         & 5.69 & -8.97  & 0.0224   & 0.0081  & 0.149 & 6.933 & 1.34\\
FSUGold2      & 5.59 & -8.88   & 0.0207   & 0.0078  & 0.145 & 6.996 & 1.38 \\
BKA20         & 5.85 & -9.03 & 0.0238   & 0.0085 & 0.140  & 7.078 & 1.18\\
G3            & 5.9  & -9.22   & 0.0245   & 0.0087  & 0.141 & 7.061 & 1.19\\
NL3*          & 5.74 & -9.08  & 0.022    & 0.0082 & 0.144 & 7.012 & 1.30\\
Z27v1         & 6.95 & -10.49  & 0.0369   & 0.0110   & 0.14  & 7.078 & 0.80\\
NL3           & 5.88 & -9.17   & 0.0213   & 0.0084 & 0.144 & 7.012 & 1.21\\
TM1           & 5.85 & -8.63 & 0.025    & 0.0086 & 0.138 & 7.112 & 1.09\\
    \bottomrule    
    \end{tabular*}
\label{tlimforces}
\end{table}

After establishing the relationship between critical properties and saturation properties of cold nuclear matter, we extend these correlation to limiting properties. In Table \ref{tlimforces}, we present the values of $T_{l}$, chemical potential $\mu$ , pressure (P), gas density ($\rho_g$), liquid density ($\rho_l$), radius (R) and lifetime ($\tau$) of $^{208}$Pb nucleus for the forces considered in Table \ref{criticalparamaters}. To establish the relation of different properties we calculate the correlation  matrix for limiting properties of $^{208}$Pb nucleus, critical properties of infinite nuclear matter $T_c$ and saturation properties of cold nuclear matter. 

\begin{figure}[h]
    \centering
    \includegraphics[scale=0.35]{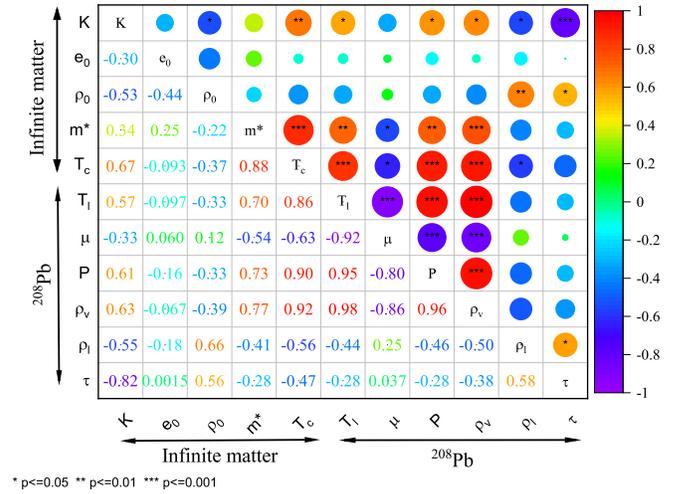}
    \caption{ The Pearson correlation matrix for the critical parameter for infinite symmetric nuclear matter, some cold nuclear matter properties and limiting properties for the $^{208}$Pb. }
    \label{tctl}
\end{figure}

Once again, the binding energy and saturation density of cold nuclear matter is weakly correlated with the limiting properties. The incompressibility  shows a weak correlation with the limiting properties which is in agreement with the analysis of  the low density behaviour of EoS. However, it is  correlated negatively with the lifetime of nucleus. This is justified as the stiff EoS corresponds to the the larger pressure, which in turn make the nucleus less stable surrounded in a nucleon gas. 
The effective mass is strongly correlated with the limiting properties. A strong correlation between T$_c$ and effective mass then suggest that the limiting properties of a nucleus essentially depend on the T$_c$ and M$^*$ of the model applied. This statement has a far reaching implication as the majority of the calculations employing statistical thermodynamics as well as compress liquid-drop model (CLDM) in astrophysical applications heavily depend on the value of T$_c$ for surface energy. Also in the microscopic calculations where the surface energy is determined using the derivative of mean-fields, effective mass plays the determining role. On the other hand, the limiting properties for $^{208}$Pb i.e. limiting temperature (MeV), chemical potential (MeV), pressure (MeV fm$^{-3}$), gas density ($fm^{-3}$), liquid density ($fm^{-3}$) and radius ($fm$) are tightly correlated. A higher  $T_{l}$  means that the chemical potential will be smaller and the equilibrium pressure and gas density will be larger.

\section{\label{conclusion} Summary and Outlook}
 
In summary, we use the effective relativistic mean-field theory (E-RMF) to analyze the thermal properties of hot nuclei. The free energy of a nucleus is estimated by using temperature and density-dependent parameters of the liquid-drop model. We parametrize the surface free energy using two approaches based on the sharp interface of the liquid-gaseous phase and the semi-classical Seyler-Blanchard interaction. The later parametrization estimates relatively stiff behavior of excitation energy, entropy, and fissility parameter. The estimations of these properties are in reasonable agreement with the available theoretical microscopic calculations and experimental observations. 

It has been observed that the thermal properties of the finite nuclear system are influenced strongly by the effective mass and critical temperature ($T_c$)  of the E-RMF parameter sets employed. A larger effective mass corresponds to the higher excitation energy, level density, limiting temperature, and limiting excitation energy. The limiting temperature also depends on the behavior of  EoS at subsaturation densities which helps to calculate the properties of surrounding nuclear gas in equilibrium with the hot nucleus. A stiff EoS at subsaturation density corresponds to the larger limiting temperature. The temperature-dependent liquid-drop fission barrier is also influenced by the $T_c$. A larger $T_c$ estimates a larger temperature where the barrier vanishes.

Finally we have performed a detailed Correlation matrix analysis to account for the large deviations in the value of critical parameters among various E-RMF sets.  The effective mass shows a strong positive correlation with the critical parameters namely ($T_c$,  $\rho_c$, $P_c$) and limiting temperature of the nucleus, which is consistent with the analytical analysis. 
The binding energy and saturation density act as independent parameters which prompts us to establish a simple multiple linear regression (MLR) between the $T_c$ and saturation properties of cold nuclear matter. Our MLR equation fits the original  $T_c$ and gives useful relationship between saturation properties and critical temperature.

The present calculations can be extended to various astrophysical problems. A similar situation is encountered in supernovae explosion and neutron star crust, where the nuclei are surrounded in a nuclear and relativistic electron gas. The model dependence can also be studied within statistical multi-fragmentation calculations. Furthermore, a comprehensive analysis is required to address the anomaly in the magnitude of the critical temperature of nuclear matter by employing the low-density correction in the EoS.

\bibliography{limtemp}

\end{document}